\documentstyle[preprint,aps]{revtex}
\input{psfig}
\def\be{\begin{equation}}
\def\ee{\end{equation}}
\begin{document}


\title{Onset of oscillatory instabilities under stochastic modulation}

\author{Fran\c{c}ois Drolet$^{1}$ and Jorge Vi\~nals$^{2}$}
\address{$^{1}$ Supercomputer Computations Research Institute, Florida State 
University, Tallahassee, Florida 32306-4052. $^{2}$ Department of Chemical
Engineering, FAMU-FSU College of Engineering, Tallahassee, Florida 32310}

\date{\today}

\maketitle

\begin{abstract}

We study the effect of external stochastic modulation on a system with O(2)
symmetry that exhibits a Hopf or oscillatory instability in the absence of
modulation. The study includes a random component in both the control parameter
of the bifurcation and in the modulation amplitude. Stability boundaries are 
computed by either solving the stationary Fokker-Planck equation in the 
vicinity of the center manifold of the underlying deterministic system whenever
possible, or by direct numerical solution otherwise. If the modulation amplitude
has a stochastic component, the primary bifurcation is
always to standing waves at a value of the control parameter that depends on
the intensity of the fluctuations. More precisely, and to contrast our results 
with the case of a deterministic periodic forcing, the onset 
of instability in the standing wave regime is shifted from its
deterministic location, and the region of primary bifurcation to
traveling waves disappears yielding instead standing waves at negative values
of the control parameter. 
 
\end{abstract}


\section{Introduction}
\label{sec:intro}

At a Hopf bifurcation in a periodically modulated system, the trivial
state loses stability to either traveling or standing waves above onset 
depending on the amplitude of the modulation $b$.
For sufficiently small modulation amplitudes, traveling waves appear at a 
fixed value of the control parameter, $a_{R}$, independent of
$b$. The threshold for standing waves, however, is a decreasing function 
of $b$. We discuss in this paper how the existence of a stochastic component in
both $a_{R}$ and $b$ affects the nature of the bifurcation, as well as the
stability boundaries of the trivial state. The calculations presented here
are not specific to a particular system, but rather are based on 
the normal form equations appropriate for a Hopf bifurcation in a system with 
O(2) symmetry when driven by a periodic force of frequency about twice the Hopf 
frequency of the unperturbed system.

Detailed studies of Hopf bifurcations have been given for a large number of
systems \cite{re:cross93}. We mention, for example,
the transition from straight rolls to Busse oscillations  
observed in Rayleigh-B\'enard convection. This instability occurs in fluids 
of low Prandtl number ($P_r$) and at sufficiently large values of the Rayleigh
number ($R$). For instance, it is observed in air ($P_r=0.71$) when $R$ 
reaches a value close to 6000. The instability manifests itself as a periodic 
transverse distortion of the rolls that propagates along their axes. Recently, 
Clever and co-workers \cite{re:clever93}
have studied the influence of a periodic modulation of the gravitational field 
on the instability. To that end, they solved numerically the time-dependent 
nonlinear equations for three-dimensional convection using 
values of $R$ above the onset of oscillatory convection. They 
varied the amplitude and frequency of the modulation, with the latter
always set to a multiple of the fundamental (unmodulated) frequency. The main
result of their study is that for moderate values of the modulation
amplitude, a transition from traveling to standing waves takes place, with
the system's frequency response being either synchronous or
sub-harmonic. The authors also found that, as the amplitude is further 
increased, this frequency locking behavior disappears and a time-dependent 
aperiodic regime sets in. 
  
The onset of convection in binary fluids also occurs through an oscillatory 
instability when the separation ratio is negative, i.e., when the temperature 
field is destabilizing whereas the composition gradient is stabilizing 
\cite{re:cross93}. Given the large difference in time scales
between energy and mass diffusion, the process is also known as
double-diffusive convection. This type of instability is commonly
observed in directional solidification experiments when a crystal which
is being grown upwards rejects a heavier solute. The effect of a periodic
modulation of the gravitational field has been addressed theoretically
by Saunders {\em et
al.} \cite{re:saunders92} for a laterally unbounded fluid layer and stress free
boundary conditions at the top and bottom of the layer. In the region of
parameters in which the bifurcation of the unmodulated system is oscillatory,
they find that below onset of the unmodulated system there exist regions of 
instability to stationary convection (either
subharmonic or synchronous with the modulation frequency) for sufficiently 
large values of the modulation amplitude. For conditions above onset of the
unmodulated system, the bifurcation is to traveling waves for arbitrarily 
small amplitudes of the modulation. These findings are in agreement with the 
general bifurcation diagram for a system of O(2) symmetry that will be 
discussed below.

Experiments on double-diffusive convection are often performed
between two conducting plates which are also made permeable so as to 
control the solute concentration at the top and bottom of the system. 
As in the Rayleigh-B\'enard case, the primary instability
leads to a pattern of rolls, but that now propagate through the system. 
Experiments performed in long narrow cells of  
annular or rectangular geometry have shown a traveling wave pattern 
that is either uniform in space or confined to a small region of the cell
\cite{re:kolodner88} \cite{re:predtechensky94}.
The influence of a periodic modulation of the
temperature gradient has been investigated  
by Rehberg et al. \cite{re:rehberg88} who studied convection 
in a water-ethanol mixture
in a small rectangular cell, starting from a uniform traveling wave pattern.
As was the case in the gravitationally modulated fluid layer, they observed 
the emergence of a standing wave structure as a periodic modulation of
sufficient amplitude was added.  

In the same paper, these authors report results from a much more 
elaborate study of the onset of electro-hydrodynamic
convection in the nematic liquid crystal Merck Phase V.
In this system, an electrostatic potential difference applied across the 
experimental cell plays a role similar to that of the temperature gradient in
thermally induced convection. 
In order to suppress charge injection processes at the electrodes,
the applied voltage is alternating at a frequency $\omega$.
As its r.m.s. value is increased, the motionless state loses stability 
to a roll pattern, the properties of which depend on the driving frequency 
$\omega$. For the Merck phase V system, steady Williams rolls emerging at low 
frequencies give way to spatially homogeneous traveling waves as the driving
frequency is increased. The authors studied the stability of these traveling 
waves against a small periodic modulation of the voltage. This perturbation, 
superimposed on top of the basic ac driving, had a small
frequency $\omega_m$ in resonance with that of the traveling waves 
(i.e., $\omega_m \approx 
2 \omega_{TW} <<w$, with $\omega_{TW}$ the frequency of the traveling waves in
the unmodulated state). As in the two cases described above,
traveling waves were found to lose stability with 
respect to standing waves as the amplitude of the modulation was 
gradually increased. A corresponding shift in the threshold was observed,
with the convecting state appearing at smaller values of the control 
parameter (i.e., the r.m.s voltage of the ac source).  

A general description of a Hopf bifurcation in a periodically modulated system
has been given by Riecke, Crawford and Knobloch \cite{re:riecke88}. 
Their analysis, which is briefly reviewed in Section \ref{sec:rie}, 
involves two complex amplitude equations governing left and 
right traveling waves emerging at a Hopf bifurcation.   
The periodic modulation, which is assumed small, provides a linear 
coupling between the two, and leads to the excitation of standing 
waves under certain conditions.
Different branches of the bifurcation diagram mark the onset of standing or 
traveling waves, and they join at a codimension-2 bifurcation point  
which has been observed in the electro-hydrodynamic convection experiments of
Rehberg et al. The model also predicts a number of secondary
instabilities which have yet to be observed experimentally.
  
The purpose of this paper is to extend the results summarized above to  
cases in which either the control parameter or the amplitude of the
modulation fluctuate randomly. Although this is a quite general question,
we are specially motivated by experiments conducted in a microgravity
environment \cite{re:walter87,re:koster90}. 
There, the effective gravitational field is known to 
fluctuate in time with the amplitude of the fluctuation being two or three
orders of magnitude larger than the residual steady gravitational field
\cite{re:alexander90}. The frequency spectrum of the residual acceleration 
field, or $g$-jitter, typically comprises periodic
components and a white noise background \cite{re:thomson97}. The physical
origin of these disturbances lies in the many mechanical processes that take
place onboard spacecraft, and their coupling to mechanical modes of the 
structure. A recent analysis of actual acceleration data taken during a Space 
Shuttle flight has shown the existence of several periodic components of
frequencies in the range of a few $Hz$, and amplitudes of the order of $10^{-3}
g_{E}$, where $g_{E}$ is the intensity of the Earth's gravitational field.
There appears to be also a white noise background with approximatively 
gaussian statistics. We attempt to present here a general framework 
within which to analyze the effects of such a residual field on an 
oscillatory instability, and therefore to provide the 
basis for future studies of specific systems. Two areas of concern include 
the appearance of undesired
instabilities of some base state caused by g-jitter, and the modification in
character and location of onset of a given instability because of the 
random component of the effective gravitational field. The specific cases of
directional solidification and double diffusive convection under reduced
gravity conditions have been reviewed in \cite{re:coriell90}. 

In the classical deterministic case, a system is said to undergo a bifurcation
when its long time behavior changes qualitatively as some
control parameter is continuously varied. Mathematically, this change 
corresponds to an exchange of stability between different 
solutions to the system's governing equation(s). The nature of the bifurcation 
depends on that of the solutions it involves:  
the saddle-node, transcritical and pitchfork bifurcations, for instance,
all involve two fixed-point solutions, while the Hopf bifurcation has
both a fixed-point and a limit cycle. 
Each one of them has an associated set of equations, known 
as its normal form, to which any specific example transforms in a 
small region around its bifurcation point. The dimension of this set is equal 
to the smallest number of equations that can still give rise to the bifurcation 
(one equation in the first three examples given above, two in the Hopf case).
In systems described by a larger number of equations than those involved
in the normal form, the governing set of equations can be reduced close
to the bifurcation point. The reduced set defines
a surface in the phase space of the original equations  known as 
the center manifold. The existence
of this surface, which has the same dimension as the normal 
form, therefore leads to a simplified formulation  
of the problem, with an underlying separation of time scales in the evolution
of variables on, and orthogonal to, the center manifold.

The effect of random fluctuations, both of internal and 
external origin, on bifurcations has been studied in considerable detail
\cite{re:graham82,re:rodriguez85,re:horsthemke83,re:pesquera85}. 
Internal fluctuations, typically of thermal origin,
enter the governing equations linearly or
\lq\lq additively", scale with the inverse of the
system's size, and lead in general to so-called imperfect bifurcations:
the bifurcation point is smeared into a small region of size proportional
to the intensity of the fluctuations. On the contrary, externally induced
fluctuations (e.g., random changes in the externally set control parameter for
the bifurcation) typically enter the governing equations nonlinearly or \lq\lq
multiplicatively", do not satisfy any a priori scaling with the size of the
system, and the bifurcation point may remain sharp,
although its position can depend on the intensity of the noise. 
For instance, Graham \cite{re:graham82} has shown that a Hopf bifurcation 
with a fluctuating control parameter exhibits a sharp rise from the trivial 
state, without any shift in the location of threshold, and a small decrease 
in the average value of
the amplitude of the new stable state compared with the deterministic case.
As will be shown below for white gaussian noise, these results also hold 
when a periodic modulation of the control parameter is
added to the system. If the intensity of this modulation is
also allowed to fluctuate however, the system's response is much more 
complex. It leads, among others, to shifts in threshold
and to excitation of standing waves in a region of parameters
in which they were previously absent. 

The remainder of this paper is structured as follows:
Section \ref{sec:rie} briefly reviews known results on the effect 
of a resonant modulation on a Hopf bifurcation in a system with 
O(2) symmetry. The stochastic extension of the analysis is given 
in Sections \ref{sec:ar} and \ref{sec:b}. The former discusses the case
of a stochastic component in the control parameter $a_R$ while the latter
addresses the case of a random forcing $b$.

\section{Hopf bifurcation under periodic modulation}
\label{sec:rie}

The results presented in sections \ref{sec:ar} and \ref{sec:b} extend the 
work of Riecke and co-workers  on Hopf bifurcations in periodically 
driven systems \cite{re:riecke88}. For completeness, we present a brief 
overview of their work here. Close to a Hopf bifurcation,
two complex amplitude equations are needed to describe the slow 
evolution of the unstable modes. Let $\Psi$ describe the state of the system.
Then, 
\be
\label{eq:state}
 \Psi = u_1(t) e^{i q z} + u_2(t) e^{i q z} + \mbox{c.c.} 
\ee
The complex amplitudes $u_1$ and $u_2$ correspond to the two eigenvalues 
$\pm i w_H$ associated with the bifurcation (with $w_H$ the Hopf frequency 
of the limit cycle), and $q$ is the characteristic  wavenumber
of the emerging structure (inversely proportional to the roll width in 
convection experiments). General equations governing the evolution of $u_1$
and $u_2$ are obtained by imposing their invariance under 
both spatial translations ($T: z \rightarrow z+d$) and spatial reflections 
($K: z \rightarrow -z$) (O(2) symmetry).  
From Eq.(\ref{eq:state}), we have $T(u_1,u_2)=(e^{i q d} u_1, e^{i q d}
u_2)$ and $K(u_1,u_2) = (u_2^*,u_1^*)$. Since the equations for $u_1$ and
$u_2$ must remain invariant under these transformations, they have the form 
\begin{eqnarray}
\label{eq:amp}
& & \partial_t u_1 = g_1 u_1 + g_2 u_2  \nonumber\\
& & \partial_t u_2 = g_2^* u_1 + g_1^* u_2,
\end{eqnarray}
where $g_1$ and $g_2$ are nonlinear functions of the invariants 
$|u_1|^2 + |u_2|^2$, $u_1 u_2^*$ and $u_1^* u_2$, and of the external
modulation $\alpha$. The analysis is further restricted to the strong 
resonance case in which the frequency of the modulation 
is almost twice the natural frequency $w_H$ of the system, or $\alpha \approx
b e^{2 i w_H t}$, with $b$ real. Letting $u_1 = \eta e^{i w_H t}$ and 
$u_2=\zeta e^{- i w_H t}$ in Eq. (\ref{eq:amp}) and dropping oscillatory terms, 
one obtains to cubic order the equations 
\be
\label{eq:eta}
 \partial_t \eta = a \eta + b \zeta + c \eta (|\eta|^2 + |\zeta|^2)
  + g \eta|\zeta|^2
\ee
and
\be
\label{eq:zeta}
  \partial_t \zeta = a^* \zeta + b \eta + c^* \zeta
    (|\eta|^2 + |\zeta|^2) + g^* \zeta |\eta|^2,
\ee
governing respectively the evolution of left ($\eta$) and right ($\zeta$)
traveling waves. 
The real part of $a$ ($a_R$) is the control parameter 
while its imaginary component ($a_i$) is the detuning of the wave from 
subharmonic resonance. 
The stability of the trivial state $\eta =\zeta = 0$ is determined 
by linearizing Eqs. (\ref{eq:eta}) and (\ref{eq:zeta}) and letting
$\eta= \bar{\eta} e^{\lambda t}$ and
$\zeta= \bar{\zeta} e^{\lambda t}$.
Solutions of that form exist provided 

\be
  \bar{\zeta}= \frac{b \bar{\eta}}{\lambda - a^*} \hspace{.8cm}
  \mbox{and} \hspace{.8cm} \lambda = a_R \pm \sqrt{b^2 - a_i^2}.
\ee
 Thus, the system undergoes either a steady bifurcation at
 $a_R = \pm \sqrt{b^2 - a_i^2} \hspace{.2cm} (b>a_i)$, or a Hopf
 bifurcation at $a_R=0 \hspace{.2cm} (b<a_i)$. The point
 $(a_R,b)=(0,a_i)$ delimiting the two corresponds
 to a codimension-two Takens-Bogdanov (TB) bifurcation point.
 
\begin{figure}
\vspace{1cm}
\psfig{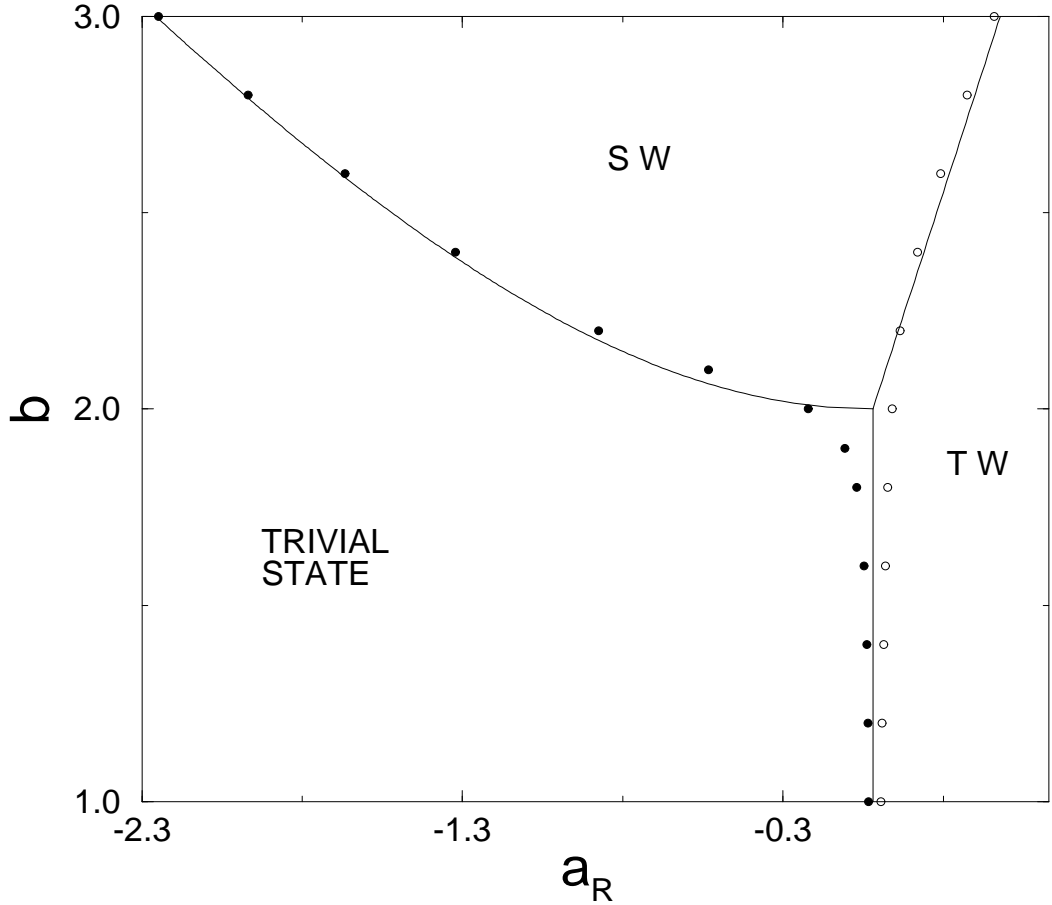}
\caption{ The solid lines are the  stability boundaries of Eqs. (\protect{\ref
{eq:x}}), (\protect{\ref{eq:y}}) and (\protect{\ref{eq:chi}}) 
 with $a_i=2, c_R=-1, c_i=2, g_R=-1 {\mbox{ and }} g_i=1$. 
\hspace{.2cm} ($\bullet$) \hspace{.05cm}: 
 onset of standing waves when fluctuations of intensity $\kappa=0.01$ are added 
to the modulation amplitude $b$;\hspace{.2cm} ($\circ$) \hspace{.05cm}: 
onset of traveling waves under the same conditions.}
\label{fig:phd}
\end{figure}

In order to study the full non-linear behavior of the system, it is
 useful to introduce the notation $\eta=x e^{i \varphi_1}$, 
 $\zeta=y e^{i \varphi_2}$, $\chi= \varphi_1 - \varphi_2$ and
 $\phi=  \varphi_1 + \varphi_2$. Then,
\be
 \label{eq:x}
 \partial_t x = a_R x + b y \cos \chi + c_R x (x^2+y^2) + g_R xy^2
\ee
\be
 \label{eq:y}
 \partial_t y = a_R y + b x \cos \chi + c_R y (x^2+y^2) + g_R yx^2  
\ee
\be
 \label{eq:chi}
 \partial_t \chi = 2 a_i + n_i (x^2 + y^2) - b \sin \chi (x^2 +y^2)/xy,
\ee
 with $n_i=2c_i + g_i$. The phase angle $\phi$ obeys the decoupled equation

\be
\label{eq:Phi}
\partial_t \phi = b \sin \chi (x^2 - y^2)/xy - g_i (x^2 - y^2).
\ee
In terms of these new variables, the state of system $\Psi$ now reads
\be
\label{eq:Psi2}
\Psi = x(t) e^{i[(\phi + \chi)/2 + w_H t + q z]} +
       y(t) e^{i[(\phi - \chi)/2 - w_H t + q z]}  + \mbox{c.c.}
\ee
Equations (\ref{eq:x}),(\ref{eq:y}) and (\ref{eq:chi}) admit two
types of stationary solutions: standing and traveling waves. For
standing waves (SW) $x=y$ and  $\partial_t \phi=0$. In that case,

 \be
   \label{eq:sw}
   x^2 = y^2 = -M \frac{1 \pm [1-N^2(a_R^2 +a_i^2-b^2)/M^2]^{1/2}}{N^2},
 \ee
with $M \equiv a_i n_i + a_R n_R, N^2 \equiv n_R^2+n_i^2$ and
$n_R \equiv 2 c_R + g_R$. For traveling waves (TW), $x \neq y$ and 
$\partial_t \phi \neq 0$. They correspond to solutions  

\be
  \label{eq:tw}
  x^2_{r,l}= -a_R [1 \pm (1- 4 \Delta^2)^{1/2}]/2 c_R,\hspace{.3cm}
  y_{r,l}=x_{l,r},
\ee
  with $\Delta^2 \equiv b^2 c_R^2/(a_R^2 g_R^2 + 4 \Omega^2)$ and
  $\Omega \equiv a_i c_R -a_r n_i/2$. These solutions
  exist as long as $\Delta^2 \le 1/4$ at which point the left and
  right traveling waves merge to form a standing wave.
  The solid lines in figure \ref{fig:phd} delimit the various regions of the
  stability diagram for the parameter set
  ($a_i=2, c_R=-1, c_i=2, g_R=-1, g_i=1$).

In summary, for small modulation amplitudes the system behaves exactly as in
the unmodulated-modulated case: traveling waves appear at onset, which is located at
$a_{R} = 0$. For modulation amplitudes larger than the detuning, standing
waves are excited instead, the threshold is at $a_{R} < 0$, and is a decreasing
function of the modulation amplitude $b$.

\section{Stochastic modulation of the control parameter}
\label{sec:ar}

We begin this section with some brief considerations about the study of 
bifurcations in a
stochastic system. As already mentioned in the introduction, we do not consider
fluctuations of internal origin (thermal fluctuations, for example), but rather
fluctuations in the externally set control parameters. The latter are not
necessarily small, typically enter the equations nonlinearly or 
\lq\lq multiplicatively", and their effect is not generally a simple smearing 
of the deterministic threshold (the so-called imperfect bifurcation in the case
of fluctuations of internal origin). Leaving aside the mathematical complexity
involved in treating but the simplest cases, there remains some discussion
in the multiplicative case
about such basic questions as the proper definition of the threshold, or the
degree of generality of the results obtained vis a vis the particular details
of the model equations or the statistical properties of the fluctuating
components. As there is no general agreement on these issues, we first
outline our underlying assumptions here.

Our definition of instability or bifurcation point follows the work 
of Graham \cite{re:graham82}, and is based on the stationary solution of the
Fokker-Planck equation for the system of interest. If the system bifurcates
from the trivial state, the solution below threshold 
is a delta function centered at zero. The onset point corresponds to the 
value of the control parameter at which additional stationary solutions of the 
Fokker-Planck equation appear with some nonzero moments. 
Other non-normalizable stationary solutions that may appear below this onset 
are not considered. Second, the dimension of our starting set of equations 
(\ref{eq:eta}) and (\ref{eq:zeta}) is larger than that of the unstable 
manifold of the deterministic case. We have adopted a center manifold reduction procedure
in the stochastic case which is analogous to the one proposed by Knobloch and 
Wiesenfeld \cite{re:knobloch83}. The stationary probability distribution 
function is assumed to factor into a contribution that depends only on the slow 
variables, and another that confines the evolution of the system to
a small region around the center manifold of the underlying deterministic
system. Numerical evidence is presented supporting such a
factorization. We note that, unless fluctuations in the direction normal to 
the center manifold can be completely neglected, such a procedure is
not equivalent to adiabatically eliminating the fast variables directly from 
the original model equations, a procedure that is standard in the study  
of deterministic bifurcations. The existence 
of a random contribution to both fast and slow time scales lies at the origin
of the difference.

We now extend the model presented in section \ref{sec:rie} to include
a random component in $a_R$.  Physically, this
corresponds to a random component in the control parameter 
of the system that has a significant frequency content at $\omega \ll 
\omega_{H}$, and a correlation time that is small in the slow time scale
emerging close to the bifurcation (inversely proportional to $a_{R}$).
Under these conditions, it is also possible to assume that the random
component is gaussian and white.  For the three examples given
in Section \ref{sec:intro}, this stochastic component reflects, 
for example,  the presence of fluctuations in either the temperature 
or gravitational field (Rayleigh-B\'enard or double-diffusive
convection) or in the applied voltage (electro-hydrodynamic convection). 
Results pertaining to the onset of standing waves are presented in Section 
\ref{sec:ar1} while the transition to traveling waves is studied in Section \ref{sec:ar2}.
All numerical simulations reported below have been performed 
using an explicit integration scheme, valid to first order in $\Delta t$ (see Appendix). 

\subsection{Bifurcation to standing waves}
\label{sec:ar1}

\begin{figure}[p]
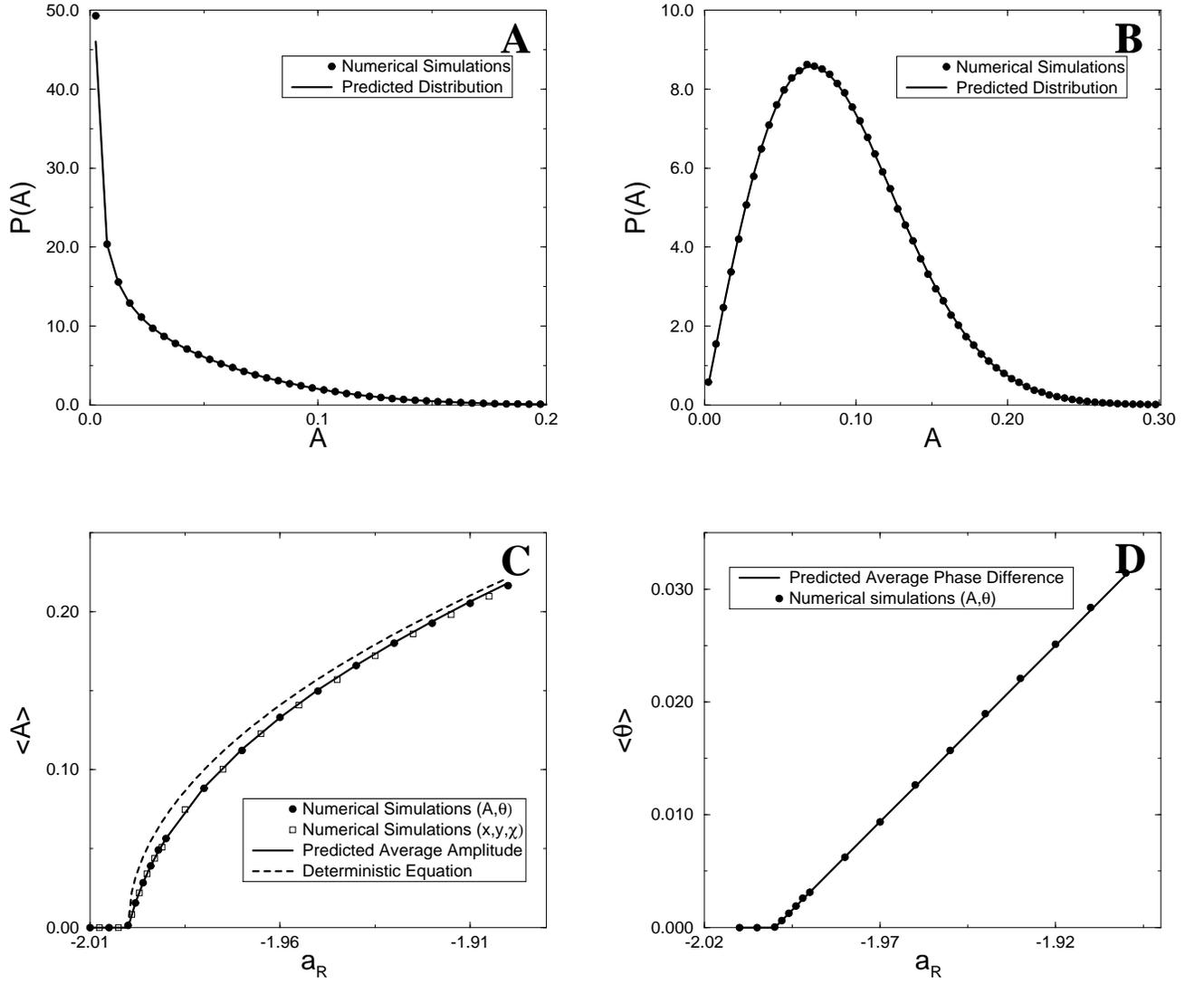

\vbox{ \hbox{
\psfig{figure=fd1.fig2a,width=3.5in}
\psfig{figure=fd1.fig2b,width=3.5in}}
       \hbox{
     \psfig{figure=fd1.fig2c,width=3.5in}
     \psfig{figure=fd1.fig2d,width=3.5in}}
      }
  \caption{$A$ and $B$: probability distribution for the 
  standing wave's amplitude above onset for the same values of the parameters as
  in Fig 1, and, $A$, $a_R  = -1.995$;
  $B$, $a_R = -1.98$. $C$ and $D$: average amplitude 
  $\langle A \rangle$ and phase difference $\langle \theta \rangle$ as 
  a function of the average control parameter $a_R$.} 
     \label{fig:noisar}
\end{figure}

Random fluctuations in the control parameter $a_R$ are introduced by letting 
$a_R \rightarrow  a_R  +\xi$ (with $a_R$ now representing  
an average value). Since it is assumed gaussian and
white, the noise $\xi(t)$ obeys the statistics 
$\langle \xi \rangle = 0$ and $\langle \xi(t) \xi(t') \rangle = 2 \kappa \delta (t-t')$, 
with $\kappa$ its intensity. 
In the region corresponding to the onset of standing waves ($b>a_i$),
it is useful to introduce the variables $A \equiv x+y$ and $Z \equiv x-y$, in
terms of which Eqs. (\ref{eq:x}) to (\ref{eq:chi}) become

\be
 \label{eq:A}
 \partial_t A = (a_R + b \cos \chi)A + \frac{c_R}{2} A (A^2+Z^2) +
   \frac{g_R}{4} A (A^2-Z^2) + \xi (t) A, 
\ee
\be
 \label{eq:Z}
 \partial_t Z = (a_R - b \cos \chi)Z + \frac{c_R}{2} Z (A^2+Z^2) - 
   \frac{g_R}{4} Z (A^2-Z^2) + \xi (t) Z, 
\ee
and
\be
 \label{eq:chi2}
 \partial_t \chi = 2 ai + \frac{n_i}{2} (A^2 + Z^2)  - 2 b \sin \chi 
  \frac{A^2 + Z^2}{A^2 - Z^2}.
\ee
Linear stability analysis performed in Section \ref{sec:rie} showed that,
in the deterministic case, standing waves appear supercritically at
$a_R= -d \equiv - \sqrt{b^2 - a_i^2}$. From the linear part of 
Eq. (\ref{eq:A}), this
implies $\cos \chi = \sqrt{1-a_i^2/b^2}$ or $\chi=\bar{\chi} \equiv
\arcsin (a_i/b)$ at the (deterministic) bifurcation point.
The trivial solution $Z=0$ remains stable above onset as the linear
coefficient $a_R - b \cos \chi$ in Eq. (\ref{eq:Z}) is negative.
Thus, an initial difference between the amplitudes of the left and right 
traveling waves rapidly decays to zero.
This is qualitatively unchanged in the stochastic equation, as the variable
$Z$ multiplies the noise $\xi(t)$ and thus suppresses the influence 
of fluctuations as it goes to zero. 
Hence, close to onset, the governing equations can be approximated by

\be
 \label{eq:redu}
 \partial_t A = (a_R + b \cos \chi)A + \frac{n_R}{4} A^3 + \xi (t) A, 
\ee
and
\be
 \label{eq:redu2}
 \partial_t \chi = 2 ai + \frac{n_i}{2} A^2  - 2 b \sin \chi.
\ee   
Furthermore, in the weak noise limit, it is 
reasonable to expect the phase
angle $\chi$ to differ only slightly from its deterministic value at onset,  
$\bar{\chi}$. Therefore, we introduce
the variable $\theta = \chi - \bar{\chi}$ and assume $\theta << 1$.
Expansion of the trigonometric functions in equations (\ref{eq:redu})
and (\ref{eq:redu2}) yields  to first order in $\theta$  
\begin{eqnarray}
 \label{eq:red}
    \partial_t \left[ \begin{array}{c} A \\ \theta
        \end{array}  \right]  & = & \left[ \begin{array}{cc}
         a_R + d & 0 \\
        0& -2 d \end{array} \right]
        \left[ \begin{array}{c} A\\ \theta \end{array} \right] +
        \left[ \begin{array}{c}
         \frac{n_R}{4} A^3 -a_i \theta A  \\
         \frac{n_i}{2} A^2 
    \end{array} \right] \\ \nonumber
   & & + \left[ \begin{array}{cc} 1 & 0 \\ 0 & 0 \end{array} \right]
   \left[ \begin{array}{c} A\\ \theta \end{array} \right] \xi (t).  
 \end{eqnarray}
Just above onset, the two eigenvalues $\lambda_1 \equiv a_R+d$ and
$\lambda_2 \equiv -2 d$ of the linearization
matrix $M \equiv \left[ \begin{array}{cc} a_R + d & 0 \\ 0& -2 d \end{array}
 \right]$
are of opposite sign. Furthermore, $\lambda_1 << 1$ while $|\lambda_2|$ is
of order unity except in the close vicinity of the codimension-two bifurcation
point (where $d \rightarrow 0$). This implies the existence of two 
different time scales in the problem, the first one of which characterizes
the rapid relaxation of the system to the center manifold $\theta_o(A)$. To
lowest order in $A$, $\theta_o(A) = (n_i/4d)A^2$, as seen by letting 
$\partial_t \theta = 0$ in Eq. (\ref{eq:red}).  
The fact that $\xi(t)$ does not appear in the equation for 
$\theta$ allows us to neglect fluctuations away from $\theta_o(A)$. The
subsequent evolution of the system is therefore confined to the center
manifold, and we look for  stationary solutions of the form 
${\cal P}(A,\theta)= P(A) \delta (\theta - \theta_o(A))$ to the Fokker-Planck
equation corresponding to Eq. (\ref{eq:red}).
Explicitly, the time-independent probability
distribution ${\cal P}(A,\theta)$ describing the statistical properties
of the system obeys the equation 

 \begin{eqnarray}
    \label{eq:fp}
   - \frac{\partial}{\partial A}
  \left\{ \left[ (a_R+d) A +  \frac{n_R}{4} A^3 - a_i \theta A + \kappa A   
     \right] {\cal P} - \kappa \frac{\partial}{\partial A} \left[ A^2 {\cal P}
    \right] \right\} \\ \nonumber  -  \frac{\partial}{\partial \theta} 
    \left[ \left( -2d \theta 
   + \frac{n_i}{2} A^2 + a_i \theta^2 \right) {\cal P} \right] = 0  .  
 \end{eqnarray}            
The fast variable is eliminated from the dynamics by integrating 
this equation
over $\theta$, with ${\cal P}(A,\theta)= P(A) \delta (\theta - \theta_o(A))$.  
The second term on the L.H.S. vanishes once the integral 
is performed, as it is proportional to ${\cal P}(A,\theta)$ evaluated 
at the limits of integration. This leaves an ordinary differential equation 
for $P(A)$
\be
   \left[ (a_R+d) A + \left(n_R - \frac{a_i n_i}{d} \right) \frac{A^3}{4} 
   - \kappa A \right] P - \kappa A^2 \frac{d}{d A} P = 0, 
\ee  
  with solution 
 \be
   \label{eq:proba}
   P(A) = {\cal N} A^{\frac{a_R+ d}{\kappa}-1}
          \exp \left[\left( n_R - \frac{a_i n_i}{d} \right)
         \frac{A^2}{8 \kappa} \right].
 \ee
 This distribution is normalizable (with ${\cal N}= 2
  [-( n_R - a_i n_i/d)/ 8 \kappa]^{\frac{ar+d}{2 \kappa}}
  / \Gamma [(a_R + d)/2 \kappa]$
 the normalization
 constant) as long as $a_R>-d$. Below that value, $P(A) = \delta (A)$,
 which implies that, just as in the deterministic case, the 
 value $a_R=-d$ marks the onset of standing waves. 
 Just above onset, the expression given in Eq. (\ref{eq:proba}) 
 exhibits a divergence at the origin (figure \ref{fig:noisar}$A$). 
 At $a_R=-d+\kappa$, this divergence transforms into a maximum 
 which moves to the right as the
 control parameter is further increased
 (figure \ref{fig:noisar}$B$).
 Both figures, corresponding to a noise intensity $\kappa=0.01$,
 show excellent agreement between predictions 
 from Eq. (\ref{eq:proba}) and the corresponding stationary distribution 
 function obtained 
 by integrating Eq. (\ref{eq:red}) numerically. The 
 simulations were performed using
 a time increment $\Delta t = 0.01$ 
 and a bin size $\Delta A = 0.005$. Initial conditions for $A$ and $\theta$ 
 were chosen randomly from a uniform distribution in the interval $[0,.05]$. Results from 500 
 independent runs were used to compute $P(A)$.
 Each run consisted of five million transient iterations after which
 a new point was added to the statistics every 500 iterations (for
 a total of 1000 points per run).

From the distribution $P(A)$, the various moments of $A$ can also 
 be determined. In particular,
 the standing wave's average amplitude is given by   
 \be
   \label{eq:avga}
   \langle A \rangle = \left[ -\left(n_R - \frac{a_i n_i}{d} \right)
  \frac{1}{8 \kappa} \right]^{-1/2} \frac{\Gamma 
  \left( \frac{a_r + d}{2 \kappa} + \frac{1}{2} \right)}{\Gamma \left( 
  \frac{a_r + d}{2 \kappa} \right)}.  
 \ee
 As shown in figure \ref{fig:noisar}$C$, results from numerical
 integration of both 
 the reduced set [Eq. (\ref{eq:red})] and the original equations for
 $x,y$ and $\chi$ are once again in excellent agreement with predictions
 from Eq. (\ref{eq:avga}). As before, the values $\Delta t= 0.01$ 
 and $\kappa=0.01$ were
 used in each of the 50 runs performed for each value of the
 control parameter. Each run consisted of 11 million iterations 
 ($10^7$ transient)
 with new points added to the statistics every 1000 time steps. 
 At any value of the control parameter $a_R$, $\langle A \rangle < 
 A_{det}$, with $A_{det}$ the amplitude of the standing wave in the
 deterministic case (figure \ref{fig:noisar}$C$).

 The statistics of the fast variable $\theta$ also 
 follow from the analysis given above. For instance, the average phase
 difference between the left and right components of the standing wave
 is given by
\begin{eqnarray}
\label{eq:lin}
   \langle \theta \rangle & = & \int_{0}^{+\infty} dA 
 \int_{-\infty}^{+\infty} d\theta \hspace{.3cm} \theta P(A)
 \delta (\theta - \theta_0(A))
  =\int_{0}^{+\infty}dA \hspace{.3cm} \theta_0(A) P(A)
  \\ \nonumber  & = & \frac{n_i}{4d}\left[ 
 \left( \frac{a_i n_i}{d} - n_R \right) \frac{1}{8 \kappa} \right]^{-1}
  \frac{\Gamma \left(\frac{a_R +d}{2 \kappa} +1\right)}{\Gamma 
  \left(\frac{a_R +d}{2 \kappa}\right)}= \frac{n_i}{d} \left[
 \frac{a_i n_i}{d} - n_R \right]^{-1} (a_R + d).  
\end{eqnarray}
Thus the average phase difference $\langle \theta \rangle$ or, equivalently, 
the average second moment $\langle A^2 \rangle$, grows linearly 
with the control parameter $a_R$. Furthermore, the slope characterizing 
this linear increase is independent of the noise intensity, so that
both $\langle \theta \rangle$ and $\langle A^2 \rangle$ assume their 
deterministic values. Figure \ref{fig:noisar}$D$  
compares predictions from Eq. (\ref{eq:lin}) with results from numerical 
simulations.  

The separation of time scales used to obtain Eqs. (\ref{eq:proba}) and (\ref{eq:avga})  
gradually disappears as the codimension-two point is approached from above
(i.e., as $|\lambda_2| = d \rightarrow 0$). Furthermore, since $\theta_o \propto 1/d$, 
fluctuations in the phase variable $\chi$ 
grow in the vicinity of the TB point, implying that
higher order terms in $\theta$ should  be kept in Eq. (\ref{eq:red}).  
Although the  predictions from Eqs. (\ref{eq:proba}) and (\ref{eq:avga}) 
fail when $b \approx a_i$, numerical integration of Eqs. (\ref{eq:x}) to (\ref{eq:chi}) 
indicates  no qualitative change in the system's behavior: the line marking the 
onset of standing waves from the trivial state remains un-shifted from its 
deterministic location, while the waves' average amplitude above onset
is comparatively smaller.

\subsection{Bifurcation to traveling waves} 
\label{sec:ar2}

The transition to traveling waves can occur 
either from the trivial state ($b < a_i$) or from a
pre-existing standing wave pattern ($b>a_i$).
The TW state is characterized by a time-dependent phase 
angle $\phi= \varphi_1 + \varphi_2$ and a finite difference in amplitude 
between the two wave components ($Z \neq 0$). Since $A$ and
$Z$ evolve over similar time scales, the governing 
equations for $x,y,$ and $\chi$ can not be simplified as in Section 
\ref{sec:ar1}. 
However, results from a numerical study indicate that 
the general conclusions of Section \ref{sec:ar1}  also hold 
in the region below the TB point.
Hence, to the accuracy of the computations, no shift was detected in the
location of onset. 
The latter was determined by computing the asymptotic amplitudes of the left and 
right traveling waves at different values of the control
parameter $a_R$. For each one of these values, the complex equations 
(\ref{eq:eta}) and (\ref{eq:zeta}) (with noise included in $a_R$) were
integrated numerically two billion times, using a time step of maximum size 
$\Delta t= 0.0005$. For all values of $b<a_i$ considered, the bifurcation was 
observed at $a_R=0 \pm 0.002$, a value consistent with its deterministic 
location ($a_R=0$). 
As in Section  \ref{sec:ar1}, a decrease in the traveling wave's
average amplitude compared to the deterministic value was also noted
above onset. 

The presence of fluctuations in the control parameter affects 
the emergence of TW above the TB point in a different way. The 
transition from SW to TW, which occurs along
the oblique line in figure \ref{fig:phd}, in the deterministic case, takes
place over a range of control parameter values when noise is added to the system. 
This smearing of the bifurcation is due to the fact that both states 
involved in the transition have associated amplitudes $x$ and $y$ which are 
non-zero.
Therefore, contrary to the primary bifurcation,
fluctuations contribute to the dynamics on   
both sides of the bifurcation point. Another way to see this is to define 
the variable $x' \equiv x - x_{det}$ (and, similarly, 
$y' \equiv y -  y_{det}$), with $x_{det}$ the deterministic value of $x$
at the bifurcation. 
The resulting equation for $x'$ involves the stochastic term $(x_{det} + x') 
\xi (t)$, in which 
the noise multiplies both the small variable $x'$ and the constant 
$x_{det}$. The second component contributes additively to the dynamics,
leading to an imperfect bifurcation.
The transition from a standing to a traveling wave state therefore 
involves intermediate values of the control parameter $a_R$ for which the
system behaves sometimes like a SW and sometimes like a TW.
Numerically, this bifurcation interval was determined 
by monitoring the temporal evolution of the quantities $Z$ and $\partial_t
\phi$ which are both zero if the pattern is a SW. For the parameter values 
given 
above and for all the driving intensities considered, the 
interval was found to include the deterministic location of onset.

\section{Stochastic variation of the modulation amplitude}
\label{sec:b}

If the random component in the externally controlled parameters has
a significant frequency content around $\omega_{H}$, the analysis
given in Section \ref{sec:ar} needs to be modified. We have first
considered the case in which the external driving is
$\alpha = \left( b + \xi(t) \right) e^{2i\omega_{H}t}$. If the correlation
time of $\xi(t)$ is large compared with $1/\omega_{H}$ but short in
the slow time scale emerging at the bifurcation, then $\xi(t)$ can 
again be assumed to be gaussian and white. A more general choice
of $\alpha$ would involve both a random amplitude and phase. In
that case, the coupling coefficient in the normal form is no longer real
and the analysis is somewhat more involved. The resulting stability
diagram is qualitatively the same than the one presented below, and
will be discussed elsewhere.

\subsection{Bifurcation to standing waves}
\label{sec:b1}

\begin{figure}[p]
\centerline{\psfig{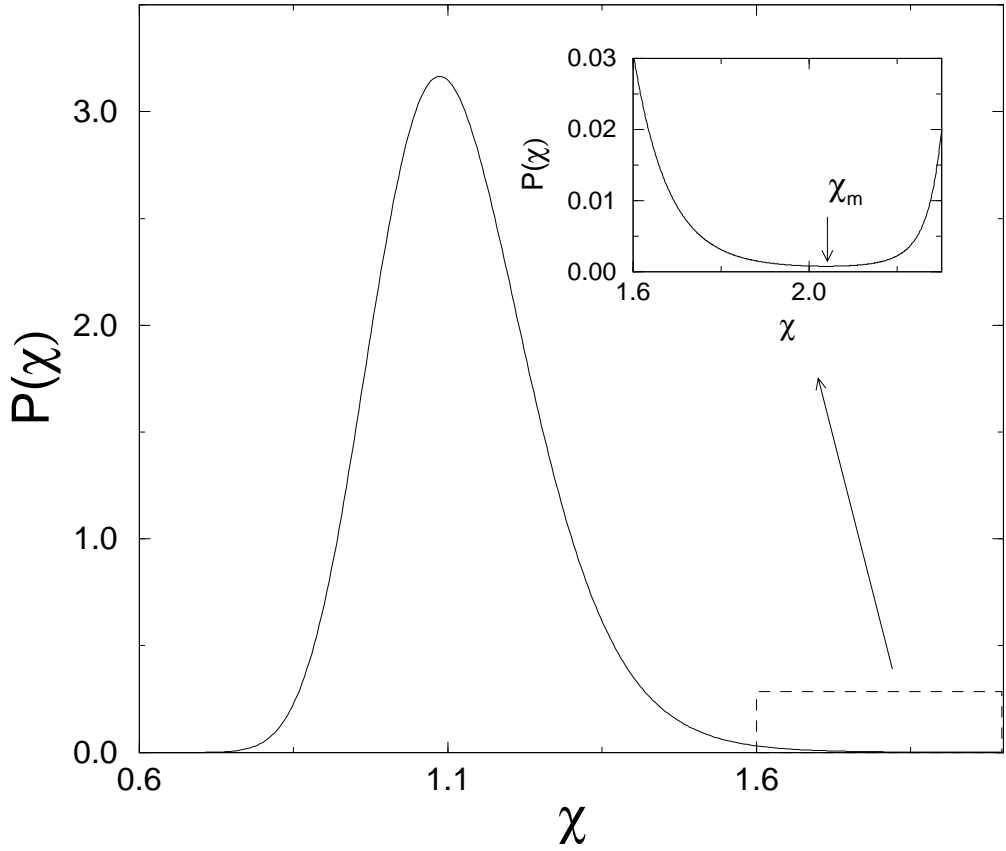}}
\caption{Probability distribution ${\cal P}(\chi)$ corresponding to the average
modulation $b=2.25$. The distribution has a minimum at some value $\chi_m$
above which it starts increasing again and eventually diverges at $\chi = \pi$
(inset).}
\label{fig:plat2}
\end{figure}

\begin{figure}
\centerline{ \psfig{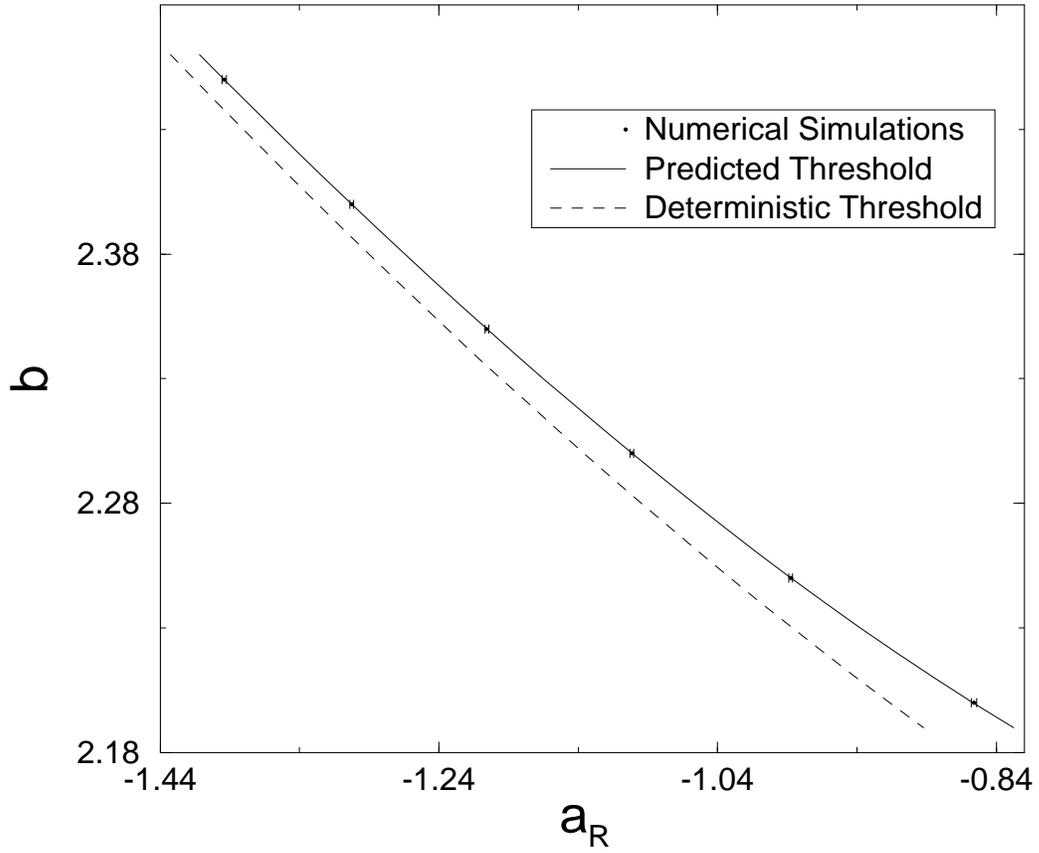} }
\caption{Location of onset in the presence of a 
fluctuating modulation amplitude $b$. For the average driving intensities 
shown, the bifurcation point is shifted to the right of its deterministic 
position.}
\label{fig:thr}
\end{figure}

\begin{figure}[p]
\centerline{\psfig{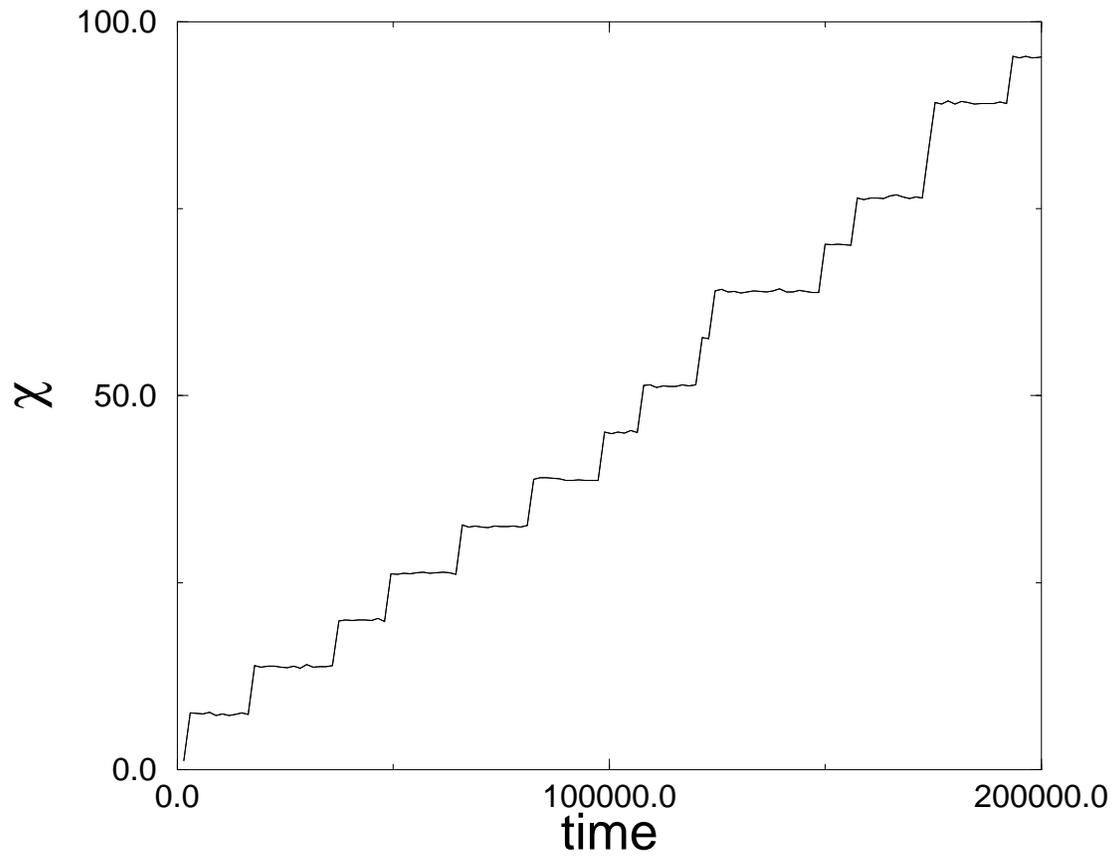}}
\caption{Temporal evolution of the phase difference $\chi$ during a typical 
run at $b=2.25$.}
\label{fig:plat}
\end{figure}

 As in Section
 \ref{sec:ar1} we first let $b= b+ \xi (t)$ and rewrite Eqs. (\ref{eq:x}) 
 to (\ref{eq:chi}) in terms of the variables $A$ and $Z$. This gives 

\be
 \label{eq:2A}
 \partial_t A = (a_R + b \cos \chi)A + \frac{c_R}{2} A (A^2+Z^2) +
   \frac{g_R}{4} A (A^2-Z^2) + \xi (t) \cos \chi A, 
\ee
\be
 \label{eq:2Z}
 \partial_t Z = (a_R - b \cos \chi)Z + \frac{c_R}{2} Z (A^2+Z^2) - 
   \frac{g_R}{4} Z (A^2-Z^2) - \xi (t) \cos \chi Z  
\ee
and
\be
 \label{eq:2chi}
 \partial_t \chi = 2 ai + \frac{n_i}{2} (A^2 + Z^2)  - 2 b \sin \chi 
  \frac{A^2 + Z^2}{A^2 - Z^2} - 2 \xi (t) \sin \chi 
  \frac{A^2 + Z^2}{A^2 - Z^2}.
\ee
 Close to onset ($a_R 
  \approx -d$, with $b>a_i$), the variable $Z$  quickly
 decays to zero and consequently drops out from the above equations.
 Thus, 

\be
  \label{eq:ab}
    \partial_t A = (a_R  + b \cos \chi)A +  n_R \frac{A^3}{4}
                 + \xi(t) \cos \chi A
\ee 
and
\be
  \label{eq:chib}
    \partial_t \chi = 2 a_i + \frac{n_i}{2} A^2 -2 b \sin \chi - 2 \xi(t) 
       \sin \chi .
\ee
 Due to the presence of non-linear functions of $A$ and $\chi$
 in the stochastic terms, the Fokker-Planck equation corresponding to 
 Eqs. (\ref{eq:ab}) and (\ref{eq:chib}) cannot be solved exactly.
 However, in the limit $A \rightarrow 0$, the 
 term $n_i A^2/2$ on the R.H.S. of Eq. (\ref{eq:chib}) can be 
 neglected, effectively decoupling equation (\ref{eq:chib}) from
 Eq. (\ref{eq:ab}). Although this 
 approximation is expected to hold only in a very small neighborhood 
 around the bifurcation point, it
 is nevertheless sufficient to determine analytically the location of
 onset, 
 which marks a transition from a 
 state with $A=0$ to one in which $\langle A \rangle$ is arbitrarily small
 (although non-zero).  
 The stationary probability distribution of the now independent
 variable $\chi$ obeys the following Fokker-Planck equation
 
\be
    \label{eq:difprob}
    \frac{d P(\chi)}{d \chi} - \left( \frac{a_i}{2 \kappa \sin^2 \chi} 
     - \frac{b}{2 \kappa \sin \chi} - \cot \chi \right) P(\chi) = 0,
\ee
 which yields

\be
    \label{eq:probchi}
      P(\chi) = {\cal N} \frac{[\tan (\chi /2)]^{-b/2 \kappa}}{\sin \chi}
                \exp \left(\frac{-a_i}{2 \kappa \tan \chi} \right).
\ee
This expression for $P(\chi)$ is plotted in figure \ref{fig:plat2}
for the average 
modulation amplitude $b=2.25$. The distribution has 
a maximum close to $\bar{\chi}=\arcsin (a_i/b)$, 
a divergence at $\chi=\pi$ and a minimum at some intermediate value
$\chi_m$ (figure \ref{fig:plat2}, inset). Except when $b\approx a_i$,
 $P(\chi_m)<<1$ so that trajectories
 are most of the time confined to the interval $[0,\chi_m]$. 
    Since the phase angle $\chi$ evolves independently  
    of $A$ and over a much shorter time scale, it effectively acts 
    in Eq. (\ref{eq:ab}) as a second noise source, with non-zero
    correlation time and non-gaussian statistics. Rewriting
    the remaining equation
    for $A$  using the variables (with zero mean) $\xi ' \equiv \cos \chi -
    \langle \cos \chi \rangle$ and $\xi '' \equiv
    \xi \cos \chi - \langle \xi \cos \chi \rangle$, we have 
  
\be
    \partial_t A = (a_R + b  \langle \cos \chi \rangle + 
    \langle \xi(t) \cos \chi \rangle) A + n_R \frac{A^3}{4} + b \xi ' A 
                 + \xi '' A, 
\ee
    which describes a pitchfork bifurcation taking place at 
     
\be
    \label{eq:newthr}
     [a_R]_{thr} =- b  \langle \cos \chi \rangle - \langle \xi \cos \chi
         \rangle.
\ee
 Using the Furutsu-Novikov theorem \cite{re:pesquera85}, 
 the second average on the R.H.S. of Eq. (\ref{eq:newthr}) simplifies to 

\be
    \langle \xi \cos \chi \rangle =  \kappa \langle \delta \cos \chi /
   \delta \xi \rangle = 2 \kappa \langle \sin^2 \chi \rangle,
\ee
   so that $[a_R]_{thr}=- b  \langle \cos \chi \rangle 
   - 2 \kappa \langle \sin^2 
   \chi \rangle$. Both averages are easily calculated from 
   the probability distribution Eq. (\ref{eq:probchi}), normalized in the
   interval $[0,\chi_m]$.
   As shown in figure \ref{fig:thr},
   excellent agreement 
   was found between predictions from Eq. (\ref{eq:newthr})
   and numerical estimates obtained directly from Eqs. (\ref{eq:x}),
(\ref{eq:y}) and (\ref{eq:chi}).
In both cases, the location of onset is shifted, indicating 
a stabilization of the trivial state. 
Simulations were performed at the two values of $a_R$ delimiting each 
error bar in figure \ref{fig:thr}. The
existence of a bifurcation within the interval  was inferred
from the large change in the asymptotic amplitudes
$x_\infty$ and $y_\infty$ noted across the interval. 
Ten runs were performed for each value of $a_R$ using 
a time step $\Delta t= 0.005$
and a total number of iterations per run $N=5 \times 10^7$.   
Although figure \ref{fig:thr} only shows results in the range $2.2<b<2.45$,
similar agreement was observed at larger values of the modulation
amplitude. 
As mentioned above however, difficulties arise when $b \approx a_i$
(i.e., close to the TB point). To understand the origin of these 
difficulties, consider the temporal evolution of the phase angle 
$\chi$ during a typical run at $b=2.25$ (figure \ref{fig:plat}). 
Long periods during which 
$\chi$ fluctuates according to the distribution Eq. (\ref{eq:probchi}) 
are followed by short intervals in which it rapidly increases by $2 \pi$.  
The existence of such steps follows from the fact that $P(\chi_m)$ is not
identically 0, allowing trajectories  in $\chi$-space to leave
the interval $[0, \chi_m]$ after a certain time. The average
time a given trajectory takes to escape is given by the
expression  \cite{re:gardiner85}

\be
 \bar T = \frac{1}{2 \kappa} \int_{\chi_o}^{\chi_m} \frac{dy}{P(y) \sin^2 y} 
  \int_0^y P(z) dz
\ee
which, evaluated for $b=2.25$, yields $\bar T = 1.30 \times 10^4$,
a result essentially independent of 
the initial condition $\chi_o$. This estimate is in good agreement with 
the value $\bar T=1.38 \times 10^4$ obtained by averaging the number
of jumps in $\chi$ 
occurring during numerical simulations. In the case just considered, 
these jumps are rare
and therefore statistically insignificant. Hence, for $b=2.25$,
Eq. (\ref{eq:probchi})
provides an accurate estimate for the averages $\langle \cos \chi \rangle$ 
and $\langle \sin^2 \chi \rangle$ which determine 
the location of onset. However, as the modulation 
amplitude $b$ is lowered, $P(\chi_m)$ increases and so does the number of 
steps in $\chi$.  The analytical approach 
developed above eventually fails and the location of onset 
must be determined numerically.  
The black dots shown in figure \ref{fig:phd}, 
mark the onset of standing waves over the entire range of $b$-values.
They indicate a qualitative change in the system's behavior near
the TB point with  
standing waves becoming stable with  
respect to the trivial state when $b \approx a_i$. 
SW are also present in the region $b<a_i$,
where they were previously unstable to traveling waves. 
Therefore, periods of rapid increase in $\chi$ 
tend to favor the formation of a SW,  
the maxima 
and minima of which get inverted with each jump in $\chi$ (see 
Eq. (\ref{eq:Psi2}) with $\chi \rightarrow \chi + 2 \pi$). 
This stabilization of the SW state can be
understood by first noting 
that close to the TB point, 
$\bar{\chi}=\arcsin{a_i/b} \approx \pi/2$. Hence, in the  deterministic
limit, the terms proportional to $\cos \chi$ in Eqs.(\ref{eq:x}) and 
(\ref{eq:y}) go to zero, 
suppressing the mutual excitation between left and right traveling
waves responsible for the emergence of  a SW pattern. 
By increasing the probability of finding values of $\chi$  
away from $\bar \chi$, the sudden jumps described above restore part 
of this constructive interaction and lead to the observed change in 
behavior.   
When the average modulation amplitude is small ($b<<1$), 
$\chi(t) \approx 2 a_i t$, as seen by letting $A,b \rightarrow 0$ in  
Eq. (\ref{eq:chib}).
From Eq. (\ref{eq:Psi2}), this result implies that, in the limit of 
small driving amplitude, 
the system is effectively oscillating
at a new frequency $w_H'= w_H +  a_i$. Furthermore, the location of
onset is at 

\be
    \label{eq:smlb}
    [a_R]_{thr}=- b  \langle \cos (2 a_i t) \rangle - 2 \kappa \langle 
    \sin^2 (2 a_i t) \rangle = - \kappa,
\ee 
  where the ensemble average has been replaced with an average over time.
  Results from simulations performed in the limit $b <<1$ agree with
  predictions from Eq. (\ref{eq:smlb}).

\subsection{Bifurcation to traveling waves}
\label{sec:b2}

The empty circles in figure (\ref{fig:phd}) mark the onset of traveling 
waves when $b$ is a fluctuating quantity. As in Section \ref{sec:ar2} 
these points were
obtained from numerical simulations of Eqs. (\ref{eq:eta}) and 
(\ref{eq:zeta}) in which changes in the amplitude
difference $Z$ and phase angle $\phi$ were used to monitor the progressive 
transition from a standing to a traveling wave state. The results
indicate a shift in the location of onset with the direction of this 
shift depending on the value of the modulation amplitude. In particular,
for values of $b$ around or below the detuning $a_i$, a delay in the onset
of traveling waves was observed. Thus, the TB point, 
which is a distinctive feature of the deterministic stability diagram,
disappears when a random component is added to the driving. 
 
\section*{Acknowledgments} This work was supported by the Microgravity Science
and Applications Division of the NASA under contract No. NAG3-1885. This work
was also supported in part by the Supercomputer Computations Research
Institute, which is partially funded by the U.S. Department of Energy,
contract No. DE-FC05-85ER25000.

\appendix
\section{}
\label{sec:app}

Numerical integration of the various stochastic differential equations 
encountered in sections \ref{sec:ar} and \ref{sec:b} was performed
using an explicit scheme valid to first order in $\Delta t$. Expressed 
in terms of the Stratonovitch calculus, the algorithm used 
\cite{re:sancho82} \cite{re:rao74} maps the
Langevin equations 

\be
\label{eq:lang}
\dot{x_i}= f_i(\{x_k(t)\}) + g_i(\{x_k(t)\}) \xi(t), 
\ee
with $\xi(t)$ gaussian white noise, to the discrete set 

\begin{eqnarray}
\label{eq:algo1}
x_i(t+\Delta t) =& &  x_i(t) + f_i(\{x_k(t)\}) \Delta t + 
  g_i(\{x_k(t)\}) \Xi(t) \\ \nonumber & & +
 \frac {1}{2} \sum_j g_j(\{x_k(t)\}) \frac{\partial g_i(\{x_k(t)\})}{
 \partial x_j(t)} 
  \Xi^2(t) +  O(\Delta t^{3/2}).
\end{eqnarray}
The random number $\Xi(t)$ is gaussian distributed, with variance 
$\langle \Xi(t)^2 \rangle = 2 \kappa \Delta t$.
As an example, we give the discretized version of Eqs.(\ref{eq:x}) to 
(\ref{eq:chi}) with 
noise included in the control parameter $a_R$:

\begin{eqnarray}
  x(t+\Delta t) =& & x(t) + \Delta t [(a_R + b \cos \chi(t)) x(t) 
 + c_R x(t) (x(t)^2 + y(t)^2) + g_R x(t) y(t)^2] \\ \nonumber & & 
   + x(t) \Xi(t) + \frac{1}{2} x(t) \Xi(t)^2, 
\end{eqnarray}

\begin{eqnarray}
  y(t+\Delta t) =& & y(t) + \Delta t [(a_R + b \cos \chi(t)) y(t) 
 + c_R y(t) (x(t)^2 + y(t)^2) + g_R y(t) x(t)^2] \\ \nonumber & & 
 + y(t) \Xi(t) + \frac{1}{2} y(t) \Xi(t)^2, \hspace{.4cm} {\mbox{and}}
\end{eqnarray}

\be
  \chi(t+\Delta t) = \chi(t) + \Delta t [ 2 a_i+ n_i (x(t)^2 + y(t)^2) 
   -b \sin \chi(t) (x(t)^2 + y(t)^2)/xy)]. 
\ee

\bibliographystyle{prsty}
\bibliography{references}
 
\newpage

\end{document}